\documentclass[aps,prd,twocolumn,superscriptaddress,amsmath,amssymb]{revtex4-1}

\usepackage{hyperref}
\usepackage{graphicx}

\newcommand{\etal}{{\it et al.}}

\newcommand{\shiki}[1]{Eq.~(\ref{#1})}
\newcommand{\zu}[1]{Fig.~{\ref{#1}}}
\newcommand{\hyou}[1]{Table~{\ref{#1}}}

\newcommand{\attn}{{\rm attn}}
\newcommand{\RSE}{{\rm RSE}}
\newcommand{\SQL}{{\rm SQL}}
\newcommand{\SRM}{{\rm SRM}}

\newcommand{\arm}{{\rm arm}}
\newcommand{\f}{{\rm f}}
\newcommand{\m}{{\rm m}}

\newcommand{\phidet}{\phi_{\det}}
\newcommand{\Msun}{M_{\odot}}
\newcommand{\fc}{{\rm fc}}
\newcommand{\rt}{{\rm rt}}
\newcommand{\dB}{{\rm dB}}

\begin{document}

\title{Prospects for improving the sensitivity of\\ the cryogenic gravitational wave detector KAGRA}

\author{Yuta~Michimura}
  \email{michimura@phys.s.u-tokyo.ac.jp}
  \affiliation{Department of Physics, University of Tokyo, Bunkyo, Tokyo 113-0033, Japan}
\author{Kentaro~Komori}
  \affiliation{LIGO Laboratory, Massachusetts Institute of Technology, Cambridge, MA 02139, USA}
\author{Yutaro~Enomoto}
  \affiliation{Department of Applied Physics, University of Tokyo, Bunkyo, Tokyo 113-8656, Japan}
\author{Koji~Nagano}
  \affiliation{JAXA Institute of Space and Astronautical Science, Sagamihara, Kanagawa 252-5210, Japan}
\author{Atsushi~Nishizawa}
  \affiliation{Research Center for the Early Universe, University of Tokyo, Bunkyo, Tokyo 113-0033, Japan}
\author{Eiichi~Hirose}
  \affiliation{inhbar, Inc., Adachi, Tokyo 121-0051, Japan}
  \affiliation{Institute for Cosmic Ray Research, University of Tokyo, Kashiwa, Chiba, 277-8582, Japan}
\author{Matteo~Leonardi}
  \affiliation{National Astronomical Observatory of Japan, Mitaka, Tokyo, 181-8588, Japan}
\author{Eleonora~Capocasa}
  \affiliation{National Astronomical Observatory of Japan, Mitaka, Tokyo, 181-8588, Japan}
\author{Naoki~Aritomi}
  \affiliation{Department of Physics, University of Tokyo, Bunkyo, Tokyo 113-0033, Japan}
\author{Yuhang~Zhao}
  \affiliation{National Astronomical Observatory of Japan, Mitaka, Tokyo, 181-8588, Japan}
  \affiliation{The Graduate University for Advanced Studies (SOKENDAI), Mitaka, Tokyo 181-8588, Japan}
\author{Raffaele~Flaminio}
  \affiliation{National Astronomical Observatory of Japan, Mitaka, Tokyo, 181-8588, Japan}
  \affiliation{Laboratoire d'Annecy de Physique des Particules (LAPP), Univ. Grenoble Alpes, Universit{\'e} Savoie Mont Blanc, CNRS/IN2P3, F-74941 Annecy, France}
\author{Takafumi~Ushiba}
  \affiliation{Institute for Cosmic Ray Research, KAGRA Observatory, University of Tokyo, Hida, Gifu 506-1205, Japan}
\author{Tomohiro~Yamada}
  \affiliation{Institute for Cosmic Ray Research, Research Center for Cosmic Neutrinos, University of Tokyo, Kashiwa, Chiba 277-8582, Japan}
\author{Li-Wei~Wei}
  \affiliation{Max Planck Institute for Gravitational Physics (Albert Einstein Institute) and Institut f{\"u}r Gravitationsphysik der Leibniz Universit{\"a}t Hannover, Callinstra{\ss}e 38, D-30167 Hannover, Germany}
\author{Hiroki~Takeda}
  \affiliation{Department of Physics, University of Tokyo, Bunkyo, Tokyo 113-0033, Japan}
\author{Satoshi~Tanioka}
  \affiliation{National Astronomical Observatory of Japan, Mitaka, Tokyo, 181-8588, Japan}
  \affiliation{The Graduate University for Advanced Studies (SOKENDAI), Mitaka, Tokyo 181-8588, Japan}
\author{Masaki~Ando}
  \affiliation{Department of Physics, University of Tokyo, Bunkyo, Tokyo 113-0033, Japan}
  \affiliation{National Astronomical Observatory of Japan, Mitaka, Tokyo, 181-8588, Japan}
  \affiliation{Research Center for the Early Universe, University of Tokyo, Bunkyo, Tokyo 113-0033, Japan}
\author{Kazuhiro~Yamamoto}
  \affiliation{Department of Physics, University of Toyama, Toyama, Toyama, 930-8555, Japan}
\author{Kazuhiro~Hayama}
  \affiliation{Department of Applied Physics, Fukuoka University, Nanakuma, Fukuoka 814-0180, Japan}
\author{Sadakazu~Haino}
  \affiliation{Institute of Physics, Academia Sinica, Nankang, Taipei 11529, Taiwan}
\author{Kentaro~Somiya}
  \affiliation{Department of Physics, Tokyo Institute of Technology, Meguro, Tokyo 152-8550, Japan}
\date{\today}

\begin{abstract}
Upgrades to improve the sensitivity of gravitational wave detectors enable more frequent detections and more precise source parameter estimation. Unlike other advanced interferometric detectors such as Advanced LIGO and Advanced Virgo, KAGRA requires different approach for the upgrade since it is the only detector which employs cryogenic cooling of the test masses. In this paper, we describe possible KAGRA upgrades with technologies focusing on different detector bands, and compare the impacts on the detection of compact binary coalescences. We show that either fivefold improvement in the $100 M_{\odot}$--$100 M_{\odot}$ binary black hole range, a factor of 1.3 improvement in the binary neutron star range, or a factor of 1.7 improvement in the sky localization of binary neutron stars is well feasible with upgrades that do not require changes in the existing cryogenic or vacuum infrastructure. We also show that twofold broadband sensitivity improvement is possible by applying multiple upgrades to the detector.
\end{abstract}


\maketitle

\section{Introduction}
The first direct detections of gravitational waves by Advanced LIGO~\cite{aLIGO} and Advanced Virgo~\cite{AdV} have opened a brand new window for us to probe our Universe. The detections of gravitational waves from binary black holes have revealed the existence of heavy stellar-mass black holes with masses exceeding $30 \Msun$~\cite{GW150914,GWTC1}, and their origins are still controversial~\cite{BBHorigins}. The detections also demonstrated the first tests of general relativity in the strong field regime~\cite{O1BBH,GRTestGW150914}. The first detection of gravitational waves from binary neutron stars, GW170817, was done coincidentally with three detectors, which enabled the identification of its electromagnetic counterparts and the host galaxy~\cite{GW170817,GW170817Multimessenger}. The multimessenger observations allowed the first demonstration of the gravitational-wave standard siren measurement of the Hubble constant~\cite{GW170817Hubble}. Moreover, nearly simultaneous detection of gravitational waves and the gamma-ray burst GRB170817A tightly constrained the difference between the speed of gravity and the speed of light to be less than a few $\times 10^{-15}$~\cite{GW170817andGRB}, and ruled out a large class of alternative gravity theories~\cite{GW170817andATG1,GW170817andATG2,GW170817andATG3,GW170817andATG4}. The GW170817 event also provided a unique opportunity to probe a neutron star equation of state and to measure the radii of neutron stars~\cite{GW170817andEOS}.

Improving the sensitivity of gravitational wave detectors enables more frequent detections and more precise source parameter estimation. Therefore, upgrades to the detectors are critical for enriching our understanding of the origin of the sources and for more precise tests of gravity theories and astronomical observations. The sensitivity improvement also enhances the detection probability of gravitational waves from rare events such as merging intermediate mass black holes~\cite{GWIMBH} or core-collapse supernovae~\cite{Fryer2011,Kotake2013}.

To this end, there have been extensive studies to reduce the fundamental noises of ground-based interferometric gravitational wave detectors, such as seismic noise, thermal noise and quantum noise~\cite{RanaRMP}. As for Advanced LIGO, there is an ongoing effort to upgrade detectors to A+~\cite{MillerAplus,AplusDesign}. The designed sensitivity of A+ is improved over that of Advanced LIGO by roughly a factor of two, which is planned to be realized by twofold reduction in the coating thermal noise from fourfold reduction in the loss angle of the coating material, and broadband quantum noise reduction. Despite a variety of studies to reduce the coating loss angle, there have been no decisive results as of yet to realize the broadband coating thermal noise reduction. The quantum noise reduction at high frequency is assumed to be 6~dB, and the broadband reduction is planned to be realized by injecting frequency dependent squeezed vacuum using a 300-m filter cavity~\cite{KimbleFilterCavity,NAOJFilterCavity,MITFilterCavity}. There is also AdV+ project to upgrade Advanced Virgo by roughly a factor of two in the designed sensitivity~\cite{JeromeAdVplus}. AdV+ is planned to be realized by increasing the mass and the diameter of the test mass mirrors from 42~kg to 105~kg and 35~cm to 55~cm, respectively, which allows the use of larger beams, leading to lower coating thermal noise. The broadband quantum noise reduction is also planned to be done with a 300-m filter cavity, and the reduction factor at high frequency is assumed to be 8~dB. A factor of two broadband sensitivity improvement leads to eightfold increase in the detection rate, and twofold reduction in the parameter estimation error. The O5 observing run with A+ and AdV+ detectors is planned to start from late 2024 or early 2025~\cite{ObservationScenario}.

KAGRA is another interferometric gravitational wave detector, which is currently being developed at the Kamioka mine in Japan~\cite{SomiyaKAGRA,AsoKAGRA,bKAGRAPhase1,KAGRAPTEP01}. KAGRA joining the global network of gravitational wave observations will help sky localization of the source from triangulation and also help resolving the distance and binary inclination degeneracy from polarization measurements~\cite{ObservationScenario}. It also increases the duty factor of multiple detectors online, and increases the likelihood of simultaneous detections with multiple detectors. Moreover, test of alternative gravity theories through additional polarization modes will enter a new regime with addition of KAGRA as a fourth detector~\cite{TakedaPolarization}. Although detailed studies on how to upgrade KAGRA is still underway, the KAGRA project is aiming for the observing run with upgraded KAGRA along with O5 observing run of A+ and AdV+.

Upgrading KAGRA will require different strategy compared with Advanced LIGO and Advanced Virgo, since KAGRA took significantly different approach for coating thermal noise reduction. While Advanced LIGO and Advanced Virgo use large fused silica mirrors at room temperature to increase the size of the beam, KAGRA plans to cool the sapphire mirrors to cryogenic temperatures. In KAGRA, extraction of the heat from the laser beam impinging on the test masses is done by the sapphire fibers suspending the test masses. Therefore, it is necessary to use thicker and shorter fibers to extract more heat, when higher laser power in injected to the test masses. While higher laser power is required to reduce quantum noise at high frequencies, thick and short fibers to extract heat increases suspension thermal noise~\cite{SomiyaKAGRA,PSOKAGRA}. Therefore, upgrading KAGRA will not be as straightforward as room temperature interferometers. This could also be an issue for future cryogenic gravitational wave detectors, such as Einstein Telescope~\cite{ET} or Cosmic Explorer~\cite{CE}.

In this paper, we present the possibility of upgrading KAGRA using established technologies which do not require changes in the existing cryogenic or vacuum infrastructure. More specifically, we discuss the impact on the sensitivity by changing the laser power, increasing the mirror mass, or injecting frequency dependent squeezing. These changes give sensitivity improvement at different detector bands, and give impacts on the detection of compact binary coalescences at different masses. We then show that twofold broadband sensitivity improvement is feasible by combining these changes.

\section{Upgrading KAGRA}
\begin{table*}
\caption{\label{KAGRAplusParams} Interferometer parameter values, inspiral ranges and median of sky localization error for GW170817-like binary for possible KAGRA upgrades. Default values~\cite{KAGRAPTEP01,LatestSensitivity} are also shown as a reference. Inspiral ranges and sky localization errors in bold are the objective function values used for the sensitivity optimization. LF: low frequency, HF: high frequency, FD SQZ: frequency dependent squeezing, FC: with 30-m filter cavity, SRC: signal recycling cavity, SRM: signal recycling mirror, BS: beam splitter.}
\begin{ruledtabular}
\begin{tabular}{llcccccc}
                       &                         & Default & LF  & HF    & Larger mirror & FD SQZ & Combined \\
\hline
SRC detuning angle (deg)& $\phidet$              & 3.5   & 28.5  & 0.1   & 3.5   & 0.2   & 0.3   \\
Homodyne angle (deg)   & $\zeta$                 & 135.1 & 133.6 & 97.1  & 123.2 & 93.1  & 93.0  \\
Mirror temperature (K) & $T_\m$                  & 22    & 23.6  & 20.8  & 21.0  & 21.3  & 20.0  \\
SRM reflectivity (\%)  & $R_{\SRM}$              & 84.6  & 95.5  & 90.7  & 92.2  & 83.2  & 80.9  \\
Fiber length (cm)      & $l_\f$                  & 35.0  & 99.8  & 20.1  & 28.6  & 23.0  & 33.1  \\
Fiber diameter (mm)    & $d_\f$                  & 1.6   & 0.45  & 2.5   & 2.2   & 1.9   & 3.6   \\
Input power at BS (W)  & $I_0$                   & 673   & 4.5   & 3440  & 1500  & 1500  & 3470  \\
Mirror mass (kg)       & $m$                     & 22.8  & 22.8  & 22.8  & 40    & 22.8  & 100   \\
\multicolumn{2}{l}{Maximum detected squeezing (dB)} & 0  & 0     & 6.1   & 0     & 5.2 (FC)& 5.1 (FC) \\
\hline
\multicolumn{2}{l}{$100\Msun$-$100\Msun$ inspiral range (Mpc)}      & 353       & {\bf 2019} & 112 & 400   & 306  & 707   \\
\multicolumn{2}{l}{$30\Msun$-$30\Msun$ inspiral range (Mpc)}        & 1095      & 1088       & 270 & 1250  & 843  & 1687  \\
\multicolumn{2}{l}{$1.4\Msun$-$1.4\Msun$ inspiral range (Mpc)}      & {\bf 153} & 85         & 155 & {\bf 202}   & {\bf 178}   & {\bf 302}   \\
\multicolumn{2}{l}{Median sky localization error (${\rm deg^2}$)}   & 0.183     & 0.506      & {\bf 0.105} & 0.156 & 0.120 & 0.100 \\
\end{tabular}
\end{ruledtabular}
\end{table*}

\def\miniwid{0.9\hsize}
\begin{figure*}
	\begin{center}
\begin{minipage}[b]{0.49\hsize}
   \begin{center}
   \includegraphics[width=\miniwid]{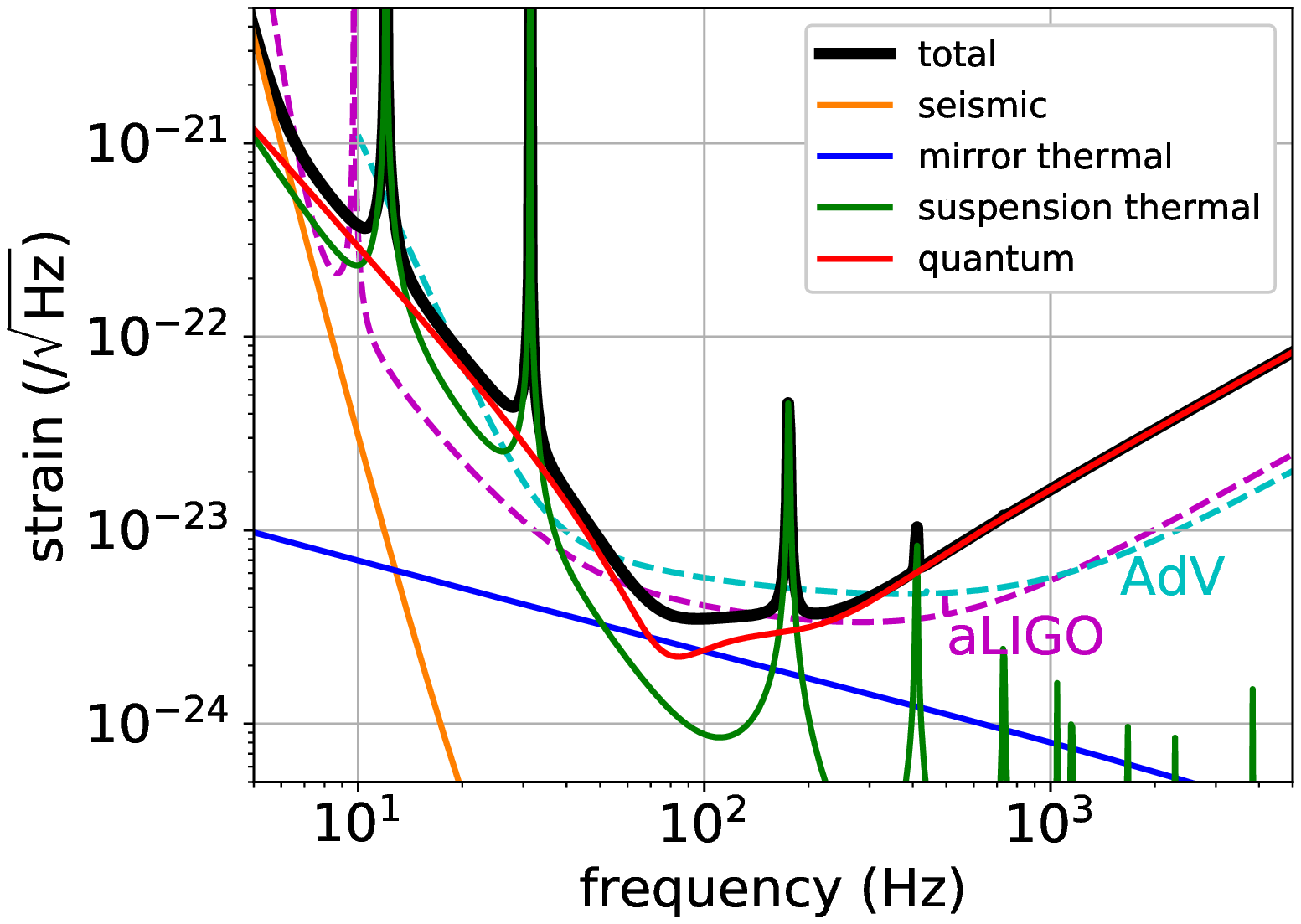} \\
   (a) Default \\
   \includegraphics[width=\miniwid]{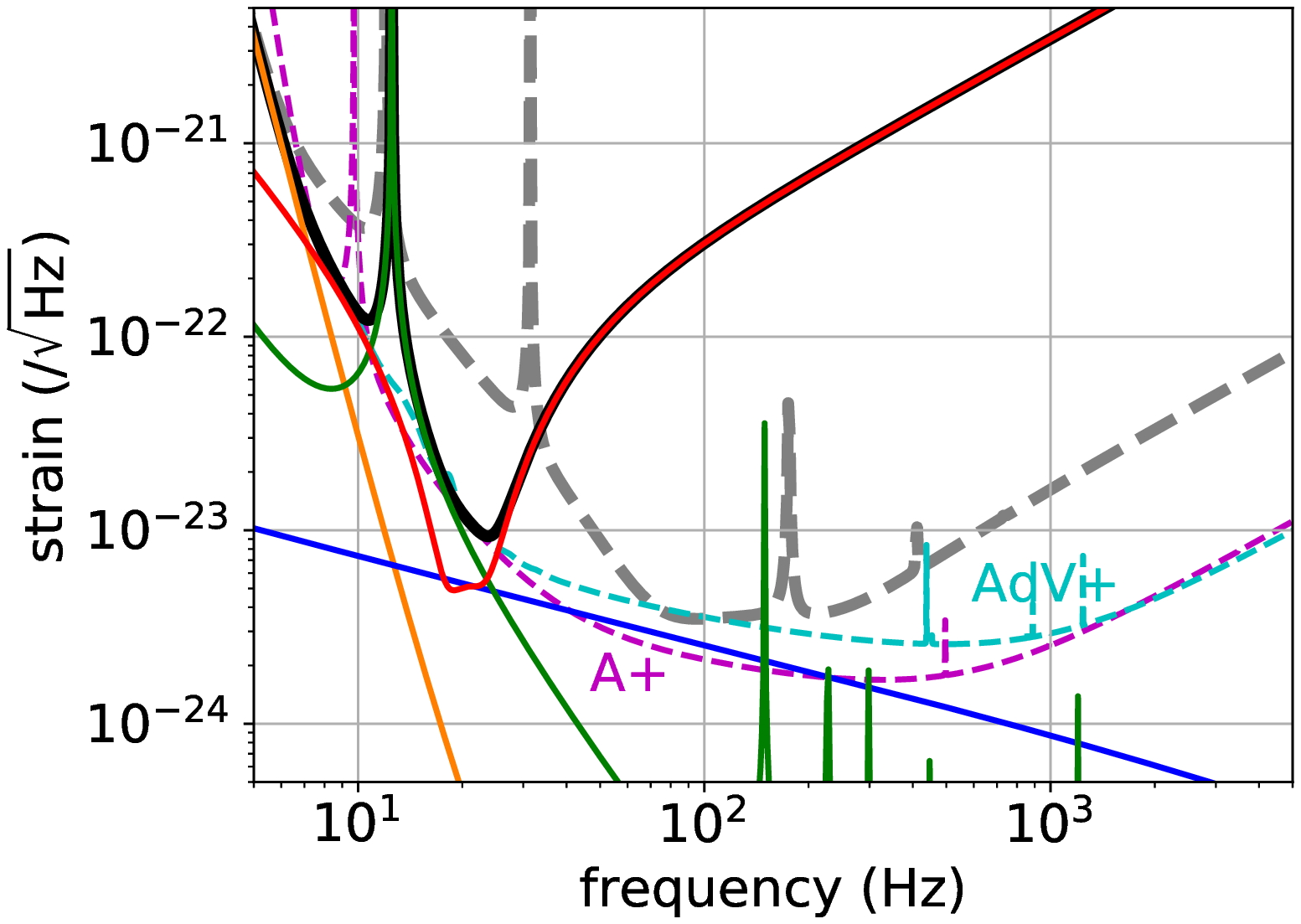} \\
   (b) Low frequency \\
   \includegraphics[width=\miniwid]{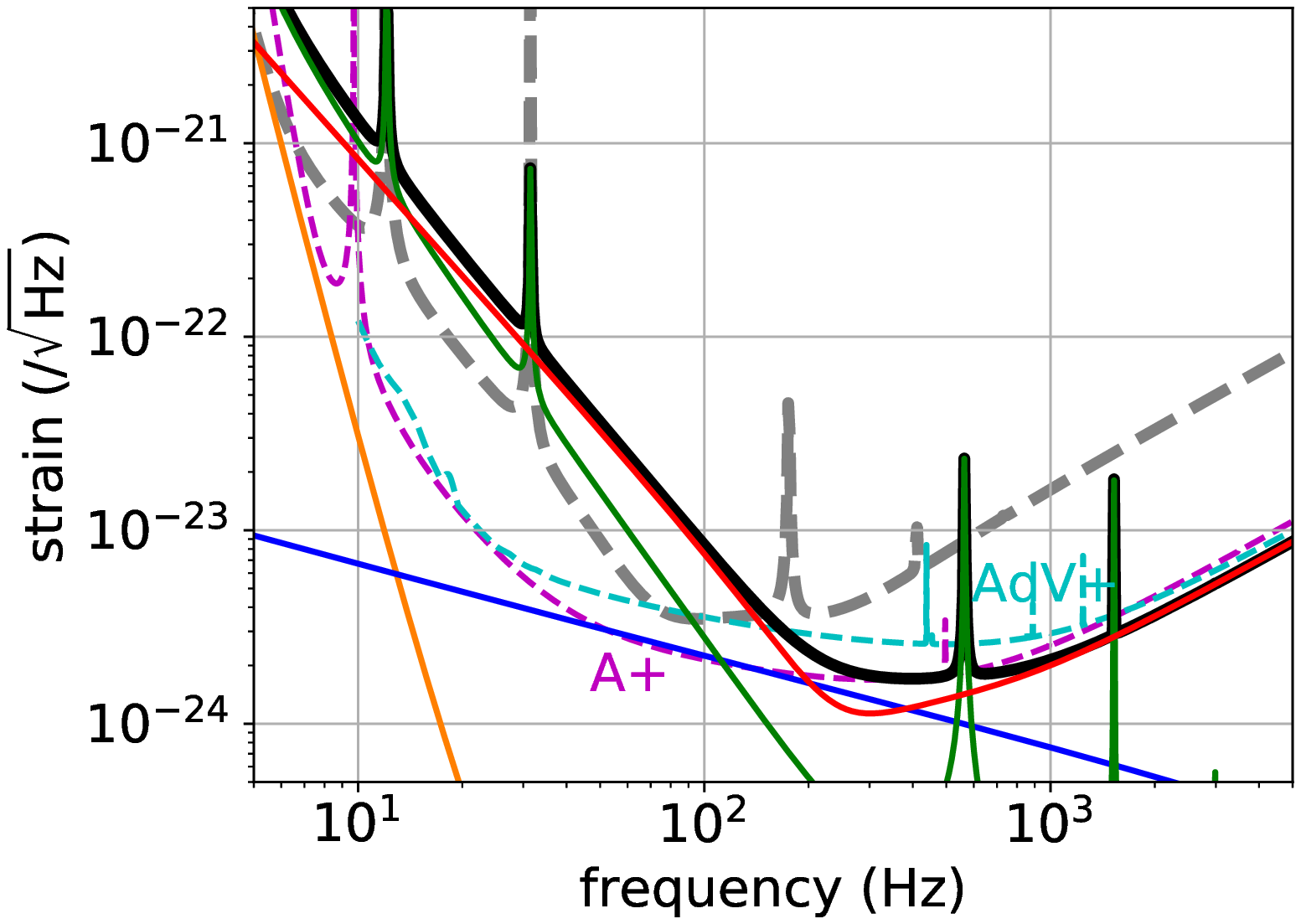} \\
   (c) High frequency
   \end{center}
\end{minipage}   
\begin{minipage}[b]{0.49\hsize}
   \begin{center}
   \includegraphics[width=\miniwid]{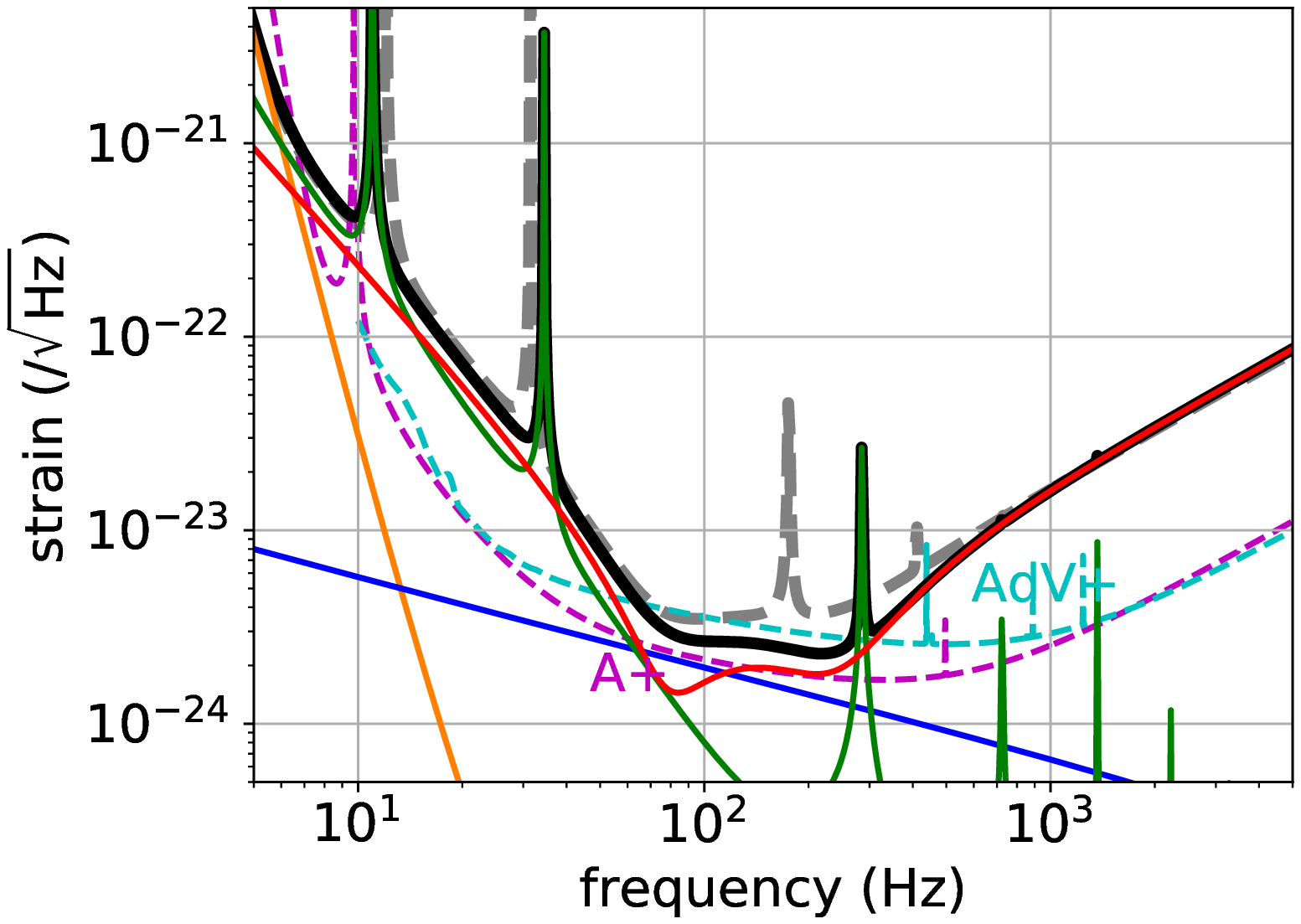} \\
   (d) Larger mirror \\
   \includegraphics[width=\miniwid]{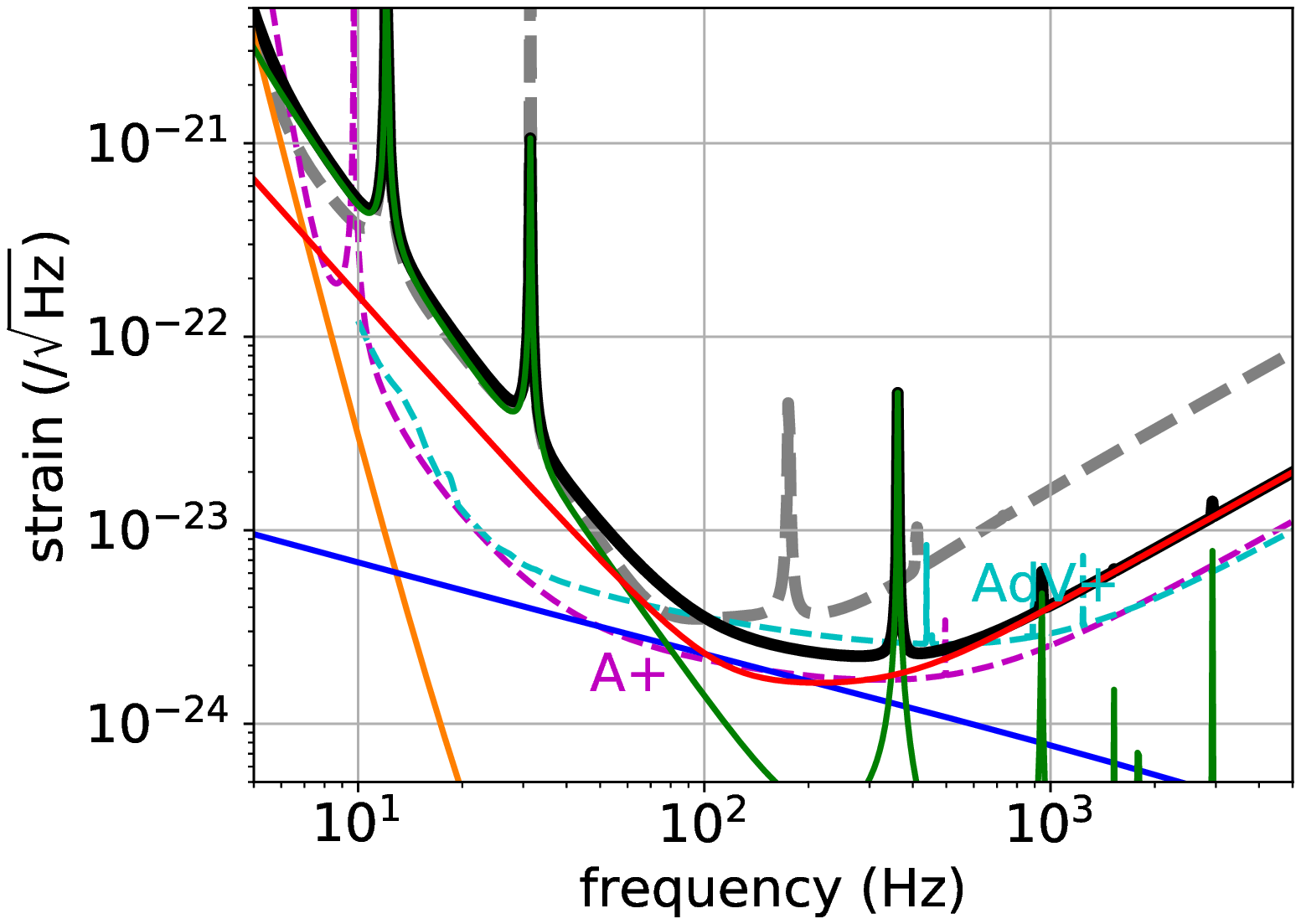} \\
   (e) Frequency dependent squeezing \\
   \includegraphics[width=\miniwid]{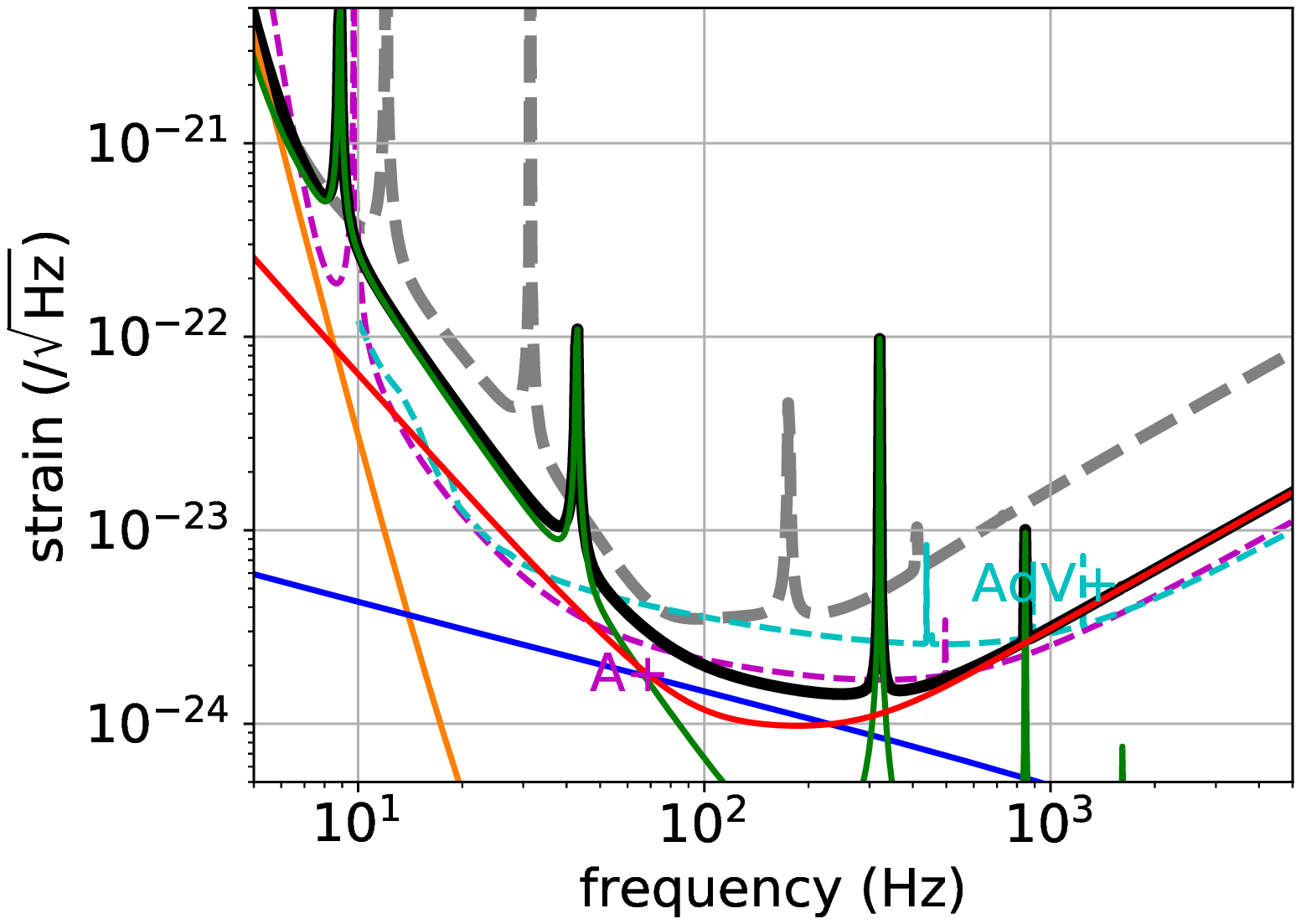} \\
   (f) Combined
   \end{center}
\end{minipage}
	\caption{\label{KAGRAplusSensitivity} Sensitivity curves for the default KAGRA and possible upgrades calculated with the parameters shown in \hyou{KAGRAplusParams}. Sensitivity curves for Advanced LIGO (aLIGO) and Advanced Virgo (AdV) are also shown for comparison in (a). For other plots, sensitivity curves for their upgrades A+ and AdV+, and default KAGRA are shown. Sensitivity curve data for Advanced LIGO and A+ are taken from Refs.~\cite{UpdatedaLIGODesign,AplusDesign}, and those for Advanced Virgo and AdV+ are extracted from Ref.~\cite{JeromeAdVplus}.}
	\end{center}
\end{figure*}

KAGRA is a dual recycled Fabry-P{\'e}rot-Michelson interferometer which employs resonant sideband extraction (RSE) technique~\cite{MizunoRSE}. KAGRA has two operation modes, such as broadband RSE and detuned RSE~\cite{SomiyaKAGRA}. In the broadband RSE case, the length of the signal recycling cavity is tuned to realize a broad detector bandwidth, while in the detuned RSE case, the length of the signal recycling cavity is detuned to maximize the binary neutron star range. The default operation mode of KAGRA is the detuned RSE, and the designed sensitivity explained in this paper is for the detuned RSE case.

The design parameters and the design sensitivity of KAGRA is shown in \hyou{KAGRAplusParams} and \zu{KAGRAplusSensitivity}(a). At low frequencies, the sensitivity is limited by suspension thermal noise and quantum radiation pressure noise. At high frequencies, the sensitivity is limited by quantum shot noise. At around 100~Hz, the sensitivity is limited by the mirror thermal noise, which mainly comes from Brownian noise of the coating.

In this section, we describe the possibility of reducing or increasing the input laser power to improve the low frequency or the high frequency sensitivity, respectively. We also describe the possibility of increasing the mirror mass to improve the sensitivity in mid-range frequencies, and the possibility of injecting frequency dependent squeezing for broad sensitivity improvement. Example parameters and sensitivities for these possible upgrades are summarized in \hyou{KAGRAplusParams} and \zu{KAGRAplusSensitivity}. The details of the sensitivity calculation and interferometer parameter tuning method are described in our previous work~\cite{PSOKAGRA}. In our parameter optimization, detuning angle of the signal recycling cavity, homodyne angle, mirror temperature, reflectivity of signal recycling mirror, length and diameter of suspension fibers, and input power at the beam splitter (BS) are tuned by particle swarm optimization.

\hyou{SearchRange} summarizes search ranges for each parameter. As for the detuning angle of the signal recycling cavity, the upper bound is set to 60~deg instead of 3.5~deg for low frequency optimization, to allow for higher detuning. The search range corresponding to the fiber diameter $d_\f$ is set using the safety factor $s_\f$ to change the range with respect to the change in the mirror mass $m$. The fiber diameter is calculated using
\begin{equation} \label{SafetyFactor}
  \pi \left( \frac{d_\f}{2} \right)^2 H_\f = s_\f \frac{m}{n_\f},
\end{equation}
where $H_\f$ is the tensile strength of the fiber, and $n_\f=4$ is the number of fibers suspending the test mass. As we have done in our previous work~\cite{PSOKAGRA}, the optimization of the input power at BS $I_0$ is done by introducing the power attenuation factor $I_\attn$,
\begin{equation} \label{AttenuationFactor}
  I_0 = I_\attn I_0^{\rm max},
\end{equation}
where $I_0^{\rm max}$ is the maximum input power allowed, calculated from the heat extraction capability of the fibers. The upper bound for the power attenuation factor was set to
\begin{equation} \label{AttenuationFactorMaximum}
  I_\attn^{\rm limit} = \frac{I_0^{\rm limit}}{I_0^{\rm max}},
\end{equation}
so that the power at the BS does not exceed $I_0^{\rm limit}$. For the high frequency upgrade plan and the plan with multiple technologies combined, we set $I_0^{\rm limit}=3500$~W, and for other upgrade plans, we set $I_0^{\rm limit}=1500$~W. Considering that the state-of-the-art high power laser source is capable of producing 400~W at the wavelength of 1064~nm~\cite{Laser400W}, we assumed such laser source is introduced for the high frequency plan and the plan with multiple technologies combined. For other plans, we assumed that the default KAGRA laser source which has the maximum output power of 180~W is used. Assuming some losses from the laser source to the interferometer input, the maximum power at BS can be determined by multiplying the power recycling gain to the maximum input laser power to the interferometer. The power recycling gain of 10 and the arm cavity finesse of 1530 are kept constant in the optimization.

To optimize the detector parameters, various astrophysical metrics have been proposed in the literature~\cite{EvanMetric,RanaMetric}. Here, we focus on the inspiral range~\cite{Finn1993} of compact binary coalescences at different masses, and the sky localization performance for binary neutron stars. To optimize the detector parameters to sky localization performance, we evaluated the error using a Fisher information matrix. The sky localization error for GW170817-like binary was calculated for 108 uniformly distributed sets of the source location and the polarization angle, and the median value was used as the objective value to be minimized. We considered the global network of two Advanced LIGO detectors at Hanford and Livingston, and Advanced Virgo at their respective designed sensitivities. Details of the calculation of objective functions are explained in Ref.~\cite{PSOKAGRA}.

\begin{table}
\caption{\label{SearchRange} The list of KAGRA detector parameters used for optimization. Their search ranges and default KAGRA values are shown. The upper bound for the detuning angle was set to 60~deg for low frequency optimization, and 3.5~deg for other optimizations. The fiber diameter is calculated using the fiber safety factor using \shiki{SafetyFactor}, and the upper bound for the power attenuation factor $I_{\attn}^{\rm limit}$ is calculated according to \shiki{AttenuationFactorMaximum}.}
\begin{ruledtabular}
\begin{tabular}{llcc}
 & & Search range & Default  \\
\hline
detuning angle (deg)   & $\phidet$   & $[0,~3.5$ or $60]$    & 3.5   \\
homodyne angle (deg)   & $\zeta$     & $[90,~180]$   & 135.1 \\
mirror temperature (K) & $T_\m$      & $[20,~30]$    & 22    \\
power attenuation      & $I_{\attn}$  & $[0.01,~I_{\attn}^{\rm limit}]$   & 1     \\
SRM reflectivity (\%)  & $R_{\SRM}$  & $[50,~100]$   & 84.6  \\
fiber length (cm)      & $l_\f$      & $[20,~100]$   & 35    \\
fiber safety factor    & $s_\f$      & $[1,~30]$     & 12.6  \\
\end{tabular}
\end{ruledtabular}
\end{table}

\subsection{Low frequency improvement}
Improving the sensitivity at the lower end of the detector band below $\sim 100$~Hz is particularly important for increasing the probability of detecting binary black holes with heavy masses. Only a
 single detection of gravitational waves from intermediate mass black holes with masses in the order of $10^2$--$10^3 \Msun$ would provide the first direct proof of their existence. Multiple detections would allow us to constrain formation scenarios of massive black holes~\cite{BBHorigins}.

The design sensitivity of KAGRA at low frequencies is limited by the quantum radiation pressure noise and the suspension thermal noise (see \zu{KAGRAplusSensitivity}(a)). These noises can be simultaneously reduced by reducing the input laser power $I_0$ since quantum radiation pressure noise scales with $\sqrt{I_0}$, and less laser power relaxes the requirement for the heat extraction through suspension fibers. The extractable heat of the fibers scales with $d_{\f}^3/l_{\f}$, when the thermal conductivity is proportional to $d_{\f}$, where $d_{\f}$ and $l_{\f}$ are the diameter and the length of the fiber, respectively. When the loss angle of the fibers are not dependent on both $d_{\f}$ and $l_{\f}$, suspension thermal noise scales with $d_{\f}/l_{\f}$. Therefore, thinner and longer fibers are necessary to reduce suspension thermal noise, if the heat extraction capability is enough.

In KAGRA, the sapphire test mass is suspended by four sapphire fibers from four sapphire blade springs attached to the intermediate mass. The intermediate mass is suspended from the upper stage by four copper beryllium (CuBe) wires. At low frequencies, not only horizontal but also vertical suspension thermal noise contributes significantly for KAGRA sensitivity~\cite{KomoriThermal}. Thermal noise peak at 12~Hz comes from resonances of sapphire blade springs, and that at 31~Hz comes from the bounce mode of CuBe wires suspending the intermediate mass. Therefore, improving the intermediate mass suspension is also required to improve the sensitivity at low frequencies.

Figure.~\ref{KAGRAplusSensitivity}(b) is an example which the sensitivity at low frequencies is improved. The mass of the intermediate mass is increased by a factor of 4 from 20.5~kg to 82~kg. The diameter and length of the CuBe wires suspending the intermediate mass is changed from 0.6~mm in diameter, 26.1~cm in length, to 0.2~mm in diameter, 78.3~cm in length. Other than the heat load from the laser power, additional heat introduced to the test mass have to be reduced from 50~mW to 3~mW. The other parameters, including the mirror mass, are kept the same as the default KAGRA parameters except for the parameters listed in \hyou{SearchRange}, which are optimized to maximize the inspiral range for 100$\Msun$-100$\Msun$ binary in the detector frame. We note that we only used inspiral signal upto the gravitational wave frequency at the innermost stable circular orbit of the Schwarzschild metric, which is 22~Hz for 100$\Msun$-100$\Msun$ binary, to calculate signal to noise ratio to focus on low frequencies here.

The optimization to produce the sensitivity curve in \zu{KAGRAplusSensitivity}(b) results in the thinnest and longest sapphire fiber within the search range. The fiber diameter of 0.45~mm is the thinnest possible diameter considering the tensile strength to suspend 22.8~kg test mass with four fibers. This low frequency upgrade also requires higher detuning of signal recycling cavity to reduce quantum noise at around 20-30~Hz, at the cost of giving up the sensitivity at high frequencies. This upgrade enables more than a fivefold improvement in the inspiral range for 100$\Msun$-100$\Msun$ binary, but the range and sky localization capability for binary neutron stars are significantly degraded.

\subsection{High frequency improvement} \label{Sec:HF}
Improving the sensitivity at high frequencies enables better sky localization of binary neutron stars, and helps electromagnetic follow-up observations. Electromagnetic follow-up observations are also crucial for more precise Hubble constant measurement. Sensitivity at the kilohertz range is also essential for the studies of post-merger signal in binary neutron star coalescences to probe neutron star physics, and for the studies of the physics behind core-collapse supernovae.

The design sensitivity of KAGRA at high frequencies is limited by the quantum shot noise. Therefore, increasing the laser power is effective since quantum shot noise scales with $1/\sqrt{I_0}$. Increasing the laser power requires thicker and shorter sapphire fibers to increase the heat extraction capability. This will result in the increase of the suspension thermal noise, but the effect to the sensitivity can be minimized when the radiation pressure noise limits the low frequency sensitivity. Another way to improve the high frequency sensitivity is to inject squeezed vacuum from the anti-symmetric port. 6~dB squeezing in the phase quadrature and 6~dB anti-squeezing in the amplitude quadrature is equivalent to increasing the input power by a factor of 4, without the necessity for changing the fiber parameters.

Figure.~\ref{KAGRAplusSensitivity}(c) is an example which the sensitivity at high frequencies is improved. Here, 10~dB of injected squeezing is assumed, with injection loss $\Lambda_{\rm inj}^2$ of 5\%.
Other optical loss parameters are kept the same as the default KAGRA values; the arm cavity round-trip loss of 100~ppm, the loss in the signal recycling cavity of 0.2~\%, and the loss in the readout $\Lambda_{\rm ro}^2$ of 10~\%.
The parameters listed in \hyou{SearchRange} are optimized to minimize the sky localization error for GW17817-like neutron star binary. For sky localization, coalescence timing measurement between the multiple detectors is important. Therefore, optimization to minimize the sky localization error is essentially the optimization to improve the sensitivity at frequencies of the coalescence. The frequency of gravitational waves at the innermost stable circular orbit of GW170817 is 1.6~kHz.

The optimization to produce the sensitivity curve in \zu{KAGRAplusSensitivity}(c) results in the thickest and shortest sapphire fiber within the search range. The mirror temperature is kept low enough not to increase the mirror thermal noise. This upgrade reduces sky localization error by a factor of 1.7 without decreasing the inspiral range for binary neutron stars. Since the low frequency sensitivity is significantly degraded, the inspiral ranges for binary black holes are degraded. Also, the optimization results in no detuning, since the detected squeezing level will be degraded with a detuned signal recycling cavity.

\subsection{Larger mirror}
Increasing the mass of the test mass generally improves the sensitivity since the suspension thermal noise and quantum radiation pressure noise scale with $m^{-3/2}$ and $m^{-1}$, respectively. The coating thermal noise can be also reduced since larger mirror allows larger beam size on the mirror. Assuming both the aspect ratio of the mirror and the ratio of the beam radius to the mirror radius to be the same, the beam radius scales with $m^{-1/3}$, and the coating thermal noise scales with $m^{-1/3}$.

Figure.~\ref{KAGRAplusSensitivity}(d) is an example sensitivity curve with the mirror mass increased from 22.8~kg to 40~kg. Considering the space available inside the current KAGRA cryostat, 40~kg mirror is the size limit without changing the design drastically. The radius and the thickness of the test mass are increased from 11~cm and 15~cm to 13~cm and 18~cm, respectively, to keep the aspect ratio. Accordingly, the beam radius at the test mass is increased from 3.5~cm to 4.2~cm. The parameters listed in \hyou{SearchRange} are optimized to maximize the inspiral range of $1.4~\Msun$-$1.4~\Msun$ binary, and all the other parameters are kept the same as default values. The coating thermal noise reduction by larger beam size is taken into account, but smaller mechanical loss of the coating material is not assumed in the sensitivity calculation to show a feasible plan. The absorption of the mirror substrate is also kept the same as the default value of 50~ppm/cm, which will result in the larger heat load for thicker test mass.

Interestingly, heavier test masses give the sensitivity improvement only at mid-frequencies where coating thermal noise dominates, when the sensitivity is optimized for binary neutron star range. This can be understood by comparing the frequency dependence of the standard quantum limit, $f^{-1}$, to that of the inspiral signal, $f^{-2/3}$. Since the standard quantum limit is steeper, it is generally better to make the standard quantum limit reaching frequency $f_\SQL$ higher to improve the inspiral range. The heavier mass requires proportionally higher laser power to keep $f_{\rm SQL}$ at the same value (see Eq.~\ref{fSQL}). In the case of KAGRA, $f_\SQL$ should be as high as possible until the quantum noise reaches the coating thermal noise. Therefore, the laser power required increases more than proportional to $m$. The input power for the default KAGRA case was limited by the heat extraction capability of sapphire fibers with the default parameters. When the fiber parameters can be tuned, the optimization results in the higher laser power with thicker fibers, and in the end the suspension thermal noise does not depend much on the mirror mass.

The optimization to produce the sensitivity curve in \zu{KAGRAplusSensitivity}(d) results in the largest input power within the search range, with shorter and thicker fiber to extract more heat. This upgrade with 40~kg test masses gives a binary neutron star range of 202~Mpc, which is a factor of 1.3 larger than the default design.

\subsection{Frequency dependent squeezing}
\begin{table}
\caption{\label{FilterCavityParams} Filter cavity (FC) parameters used for the sensitivity calculation with frequency dependent squeezing. Values in parentheses correspond to the parameters for the sensitivity curve with combined technologies. For the high frequency upgrade, the injected squeezing is also set to 10~dB, but $\Lambda_{\rm inj}^2$ is set to 5\% to take into account of having no filter cavity.}
\begin{ruledtabular}
\begin{tabular}{lcl}
Parameter & Symbol & Value  \\
\hline
FC length & $L_\fc$ &  30~m \\
FC input mirror transmissivity & $T_\fc$ & 189~ppm \\
 &  & (148)~ppm \\
FC half bandwidth & $\gamma_\fc/(2 \pi)$ &  87~Hz \\
 &  & (71)~Hz \\
FC detuning & $\Delta \omega_\fc/(2 \pi)$ & 74~Hz \\
 &  & (58)~Hz \\
FC losses & $\Lambda_\rt^2$ & 30~ppm \\
Injection losses & $\Lambda_{\rm inj}^2$ & 10\% \\
Readout losses & $\Lambda_{\rm ro}^2$ & 10\% \\
Injected squeezing & $\sigma_\dB$ & 10~dB
\end{tabular}
\end{ruledtabular}
\end{table}

Since the designed sensitivity of KAGRA is broadly limited by the quantum noise, injection of frequency dependent squeezed vacuum is a promising way improve the sensitivity. 6~dB reduction in the radiation pressure noise is equivalent to increasing the mass of the test mass by a factor of 2, without changing the suspension and coating thermal noises.

Figure.~\ref{KAGRAplusSensitivity}(e) is an example sensitivity curve with frequency dependent squeezing generated with a 30-m filter cavity. Within the KAGRA collaboration, production of the frequency dependent squeezing with a 300-m filter cavity has been recently demonstrated using the former TAMA300 interferometer infrastructure~\cite{NAOJFilterCavity,EleonoraFC1,EleonoraFC2}. As mentioned earlier, the use of 300-m filter cavities is planned for LIGO and Virgo, as longer cavities are more robust to length-dependent sources of squeezing degradation. Here, a conservative length of 30 m is used, considering the space restrictions around the signal recycling cavity and output mode cleaner chambers in KAGRA. The possibility to accommodate a longer filter cavity is still under investigation.

The filter cavity parameters used for the quantum noise calculation are summarized in \hyou{FilterCavityParams}. Using the parameters listed in \hyou{FilterCavityParams}, the half bandwidth and detuning of the filter cavity were determined with~\cite{KweeFilterCavity}
\begin{equation}
 \gamma_\fc = \sqrt{\frac{2}{(2-\epsilon)\sqrt{1-\epsilon}}} \frac{\omega_\SQL}{\sqrt{2}}
\end{equation}
and
\begin{equation}
 \Delta \omega_\fc = \sqrt{1-\epsilon} \gamma_\fc,
\end{equation}
where $\epsilon$ is the loss parameter given by
\begin{equation}
 \epsilon = \frac{4}{2 + \sqrt{2+2\sqrt{1+\left( \frac{4 L_\fc \omega_\SQL}{c \Lambda_\rt^2} \right)^4}}} .
\end{equation}
For a tuned interferometer without losses, the frequency at which the quantum noise equals the standard quantum limit $\omega_\SQL=2 \pi f_\SQL$ can be obtained by solving $\mathcal{K} = 1$, where
\begin{equation}
 \mathcal{K} = \frac{16 \pi c I_\RSE}{m \lambda L_\arm^2 \omega^2 (\gamma_\RSE^2+\omega^2)}
\end{equation}
is the optomechanical coupling constant~\cite{BuonannoChen,Kimble2001}, and is given by
\begin{equation}
 \omega_\SQL = \sqrt{\frac{-\gamma_\RSE^2+\sqrt{\gamma_\RSE^4+\frac{64 \pi c I_\RSE}{m \lambda L_\arm^2}}}{2}} . \label{fSQL}
\end{equation}
Here, $c$ is the speed of light,
\begin{equation}
 I_\RSE = \frac{1+r_\SRM}{1-r_\SRM} I_0
\end{equation}
and
\begin{equation}
 \gamma_\RSE = \frac{1+r_\SRM}{1-r_\SRM} \gamma_\arm . \label{RSEbandwidth}
\end{equation}
In the equations above, $r_\SRM=\sqrt{R_\SRM}$ and $t_\SRM$ are the amplitude reflectivity and transmissivity of the SRM, respectively, $\lambda$ is the laser wavelength, $L_\arm$ is the arm length, and $\gamma_\arm$ is the arm cavity half bandwidth. The input mirror transmissivity of the filter cavity is tuned by
\begin{equation}
 T_\fc = \frac{4 L_\fc \gamma_\fc}{c} - \Lambda_\rt^2 ,
\end{equation}
to obtain the required filter cavity bandwidth.

In the sensitivity optimization, the parameters listed in \hyou{SearchRange} are optimized to maximize the inspiral range of $1.4~\Msun$-$1.4~\Msun$ binary. All the other parameters, including the loss parameters explained in Sec.~\ref{Sec:HF} are kept the same as the default KAGRA values. We note here that filter cavity bandwidth and detuning are calculated using the equations above assuming the tuned RSE case.

As discussed in the previous section, higher input power is favored as it increases the binary neutron star range, and the optimization to produce the sensitivity curve in \zu{KAGRAplusSensitivity}(e) results in the largest input power within the search range. This requires shorter and thicker fiber to extract more heat, and results in a slightly worse suspension thermal noise. Therefore, the implementation of frequency dependent squeezing alone produces a sensitivity improvement only at high frequencies. The binary neutron star range for this example upgrade is 179~Mpc, which is a factor of 1.2 larger than the default design.

\subsection{Combining multiple technologies}
As we have shown so far, applying one of the upgrade technologies give sensitivity improvement only at certain frequencies. Combining multiple upgrades is necessary for a broadband sensitivity improvement.

The sensitivity curve in \zu{KAGRAplusSensitivity}(f) is an example with 100~kg test mass and frequency dependent squeezing with a 30-m filter cavity. The test mass of 100~kg, with the radius of 18~cm and the thickness of 25~cm, was assumed since the research and development for fabricating such sapphire mirrors is ongoing and would be feasible in the next few years. As for the filter cavity, parameters same as the ones used in the previous section was used, except for the filter cavity bandwidth and detuning to take into account the different $\omega_\SQL$ (see \hyou{FilterCavityParams}).

In the sensitivity optimization, the parameters listed in \hyou{SearchRange} are optimized to maximize the inspiral range of $1.4~\Msun$-$1.4~\Msun$ binary. The optimization to produce the sensitivity curve in \zu{KAGRAplusSensitivity}(f) results in the largest input power within the search range. The situation is similar to frequency dependent squeezing only plan, but because of larger test mass, suspension and coating thermal noises are also reduced. In total, broadband sensitivity improvement by a factor of two can be realized and the binary neutron star range will be 302~Mpc.

\section{Discussion}
\begin{figure}
\begin{center}
\includegraphics[width=\hsize]{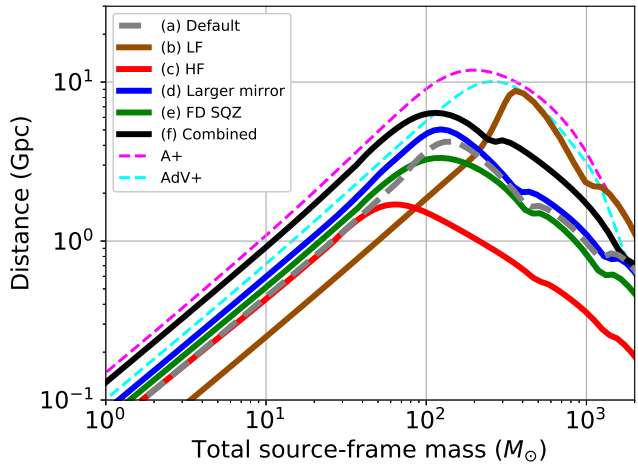}
\end{center}
\caption{\label{IMRrange} The detection ranges of possible KAGRA upgrades for non-spinning equal-mass binaries. The redshift corrected, sky averaged distance at which gravitational waves can be detected with signal-to-noise ratio of more than 8 is shown for each upgrade. LF: low frequency, HF: high frequency, FD SQZ: frequency dependent squeezing.}
\end{figure}

In the previous section, we have shown that different upgrade technologies can enhance the sensitivity at different detector bands. In this section, we briefly summarize the impact on gravitational wave science for each upgrade, and discuss the technological feasibility.

Figure.~\ref{IMRrange} shows the detection ranges of possible KAGRA upgrades discussed in the previous section. The sky averaged inspiral range using the inspiral-merger-ringdown waveform compiled in Ref.~\cite{Ajith2011} was used for calculating the detection ranges. Non-spinning equal-mass binaries are considered and the threshold for the signal-to-noise ratio is set to 8, following the convention in the gravitational wave community~\cite{Finn1993}. The mass in the x-axis is corrected to source-frame mass taking into account of the redshift.

Among four upgrade examples with a single upgrade technology, the low frequency plan has the largest detection range above $\sim 200 \Msun$ in total mass, whereas the larger mirror plan has the largest detection range for smaller masses. The detection range for the high frequency plan is the smallest, but gives the smallest sky localization error for binary neutron stars, as shown in \hyou{KAGRAplusParams}. The authors in Ref.~\cite{Uchikata2020} have shown that the frequency dependent squeezing plan, which realizes relatively broad sensitivity improvement, is most suitable for black hole spectroscopy. Each upgrade example will give different impacts on gravitational wave science. Science cases for each example are discussed in detail in Ref.~\cite{KAGRAPTEP06}.

From the technical feasibility point of view, there are some uncertainties for each upgrade. At the low frequencies, there is a variety of excess noises other than fundamental noises discussed in this paper, such as controls noise and scattered light noise~\cite{AsoKAGRA,KAGRAActuator,Den2016}. As for the plans with higher laser power, mitigation of additional effects such as thermal lensing and parametric instability~\cite{PIBraginsky,PI2015} should be considered. Thermal lensing effect in cryogenic sapphire mirrors is smaller than room temperature fused silica mirrors by orders of magnitude~\cite{ThermalLens,ThermalLensBook}, and it is unlikely to be an issue, but the effect in the fused silica BS needs to be investigated. Probability of having parametric instability with sapphire mirrors are also smaller than fused silica since number density of elastic modes are smaller, owing to the larger Young's modulus~\cite{PI2008}. However, number of unstable modes will be larger with higher power and larger mirrors.

Also, fabrication of larger sapphire mirrors would be feasible in the next few years, but it remains uncertain at this point whether larger mirror with required quality is feasible or not. Recently, in KAGRA, it is found that nonuniformity of the transmission map and inhomogeneous birefringence of the input test masses were larger than the expectation, partly due to the use of light with wrong polarization to measure the transmission map to correct the map~\cite{SomiyaTWE,HiroseMirror,KAGRAPTEP01}. These issues need to be investigated further for realizing the upgrade of KAGRA.

What will be the best choice for the next step to upgrade KAGRA depends on the technology developments discussed above, and scientific targets KAGRA will focus on. It also depends on the budget and schedule restrictions, as well as the sensitivity of other detectors. Detailed planning of the upgrade is underway within the KAGRA collaboration.

We remind here that conservative estimate in the coating thermal noise was used in the sensitivity calculation done in this work, since no coating improvements are assumed. Improvements in the heat conductivity of the sapphire suspension, and reduction of the heat absorption in both coating and sapphire substrate are also not taken into account. Developments in such technologies will bring us further enhancement of the sensitivity.

As for the interferometer parameter optimization, we only used seven parameters listed in \hyou{SearchRange}, and did not changed the arm cavity finesse or the signal recycling cavity length. The study in Ref.~\cite{Somiya2020} shows that quantum noise reduction at high frequencies is possible with higher arm cavity finesse and longer signal recycling cavity. As indicated by \shiki{RSEbandwidth}, the same quantum noise can be realized with higher arm cavity finesse and higher signal recycling mirror reflectivity, when the circulating power in the arm cavity is kept constant. This is effective for reducing the power injected to the input test mass, and therefore reducing the heat absorption.

Lastly, we note that the temperature, mirror size and suspension design of the input and end test masses are assumed to be the same. In the case of KAGRA, radius of curvatures of input and end test masses are also the same, and the beam size on the mirrors are the same, due to time and cost constraints to polish the mirrors~\cite{AsoKAGRA}. Considering that the coating thickness and the heat load of the two test masses are different, further optimization of thermal noise would be possible, if we adopt asymmetric cavity design. Optimization of the interferometer configuration with more parameters will be covered in the future work.

\section{Conclusions}
In this work, we have discussed the effect of different upgrade technologies to enhance the sensitivity of the cryogenic gravitational wave detector KAGRA. We have shown that shifting the detector band to either lower or higher frequencies is possible by changing the input power and modifying the sapphire suspension accordingly. We find that the use of larger test mass and the injection of frequency dependent squeezed vacuum will result in the sensitivity improvement mainly at mid frequencies and high frequencies, respectively, when the sensitivity is optimized for the binary neutron star range. It was demonstrated that either fivefold improvement in $100\Msun$--$100\Msun$ binary black hole range, a factor of 1.3 improvement in binary neutron star range, or a factor of 1.7 improvement in the sky localization of binary neutron stars is feasible with existing infrastructure and existing technologies. For broadband sensitivity improvement, it is necessary to apply multiple upgrades. We have shown that twofold broadband improvement is possible with 100~kg test mass, higher laser power, and the injection of frequency dependent squeezed vacuum.

Future gravitational wave detectors also plan to cool the test masses to cryogenic temperatures. As gravitational wave astronomy diversifies, future detectors would require the detector band different from that of current generation detectors, to focus on different astrophysical targets. Our study paves the way to designing the sensitivity of cryogenic gravitational wave detectors at various detector bands.

\section*{Acknowledgements}
We are grateful to J{\'e}r{\^o}me Degallaix, Matt Evans, and Stefan W. Ballmer for invaluable inputs and fruitful discussions.
This work was supported by JSPS KAKENHI Grant No. 18H01224, 18K18763, and 201960096, JST CREST Grant No. JPMJCR1873, and MEXT Q-LEAP Grant No. JPMXS0118070351.

The KAGRA project is supported by MEXT, JSPS Leading-edge Research Infrastructure Program, JSPS Grant-in-Aid for Specially Promoted Research 26000005, JSPS Grant-in-Aid for Scientific Research on Innovative Areas 2905: JP17H06358, JP17H06361 and JP17H06364, JSPS Core-to-Core Program A. Advanced Research Networks, JSPS Grant-in-Aid for Scientific Research (S) 17H06133, the joint research program of the Institute for Cosmic Ray Research, University of Tokyo, National Research Foundation (NRF) and Computing Infrastructure Project of KISTI-GSDC in Korea, Academia Sinica (AS), AS Grid Center (ASGC) and the Ministry of Science and Technology (MoST) in Taiwan under grants including AS-CDA-105-M06, the LIGO project, and the Virgo project.

This paper carries JGW Document No. {JGW-P2011740}. The sensitivity curves presented in this paper are available from Ref.~\cite{JGW-T1809537}.

%

\end{document}